\documentclass[RNAAS]{aastex631}
\usepackage{graphicx} 

\usepackage{amsmath}
\usepackage{enumitem}
\let\tablenum\relax
\usepackage{siunitx}
\usepackage{subfigure}

\usepackage{listings}

\usepackage{url}
\usepackage[sort,nameinlink]{cleveref}

\def\apjs{ApJS}

\def\aap{A\&A}
\def\apjl{ApJ}

\DeclareSIUnit\arcmin{arcmin}
\def\nside{$N_\mathrm{side}$}
\def\noise{\epsilon}

\makeatletter
\renewcommand\@makefntext[1]{\leftskip=0em\hskip0em\rightskip=0.8em\@makefnmark#1}
\makeatother

\begin{document}
\count\footins = 1000 
\title{Methods for CMB map analysis}

\author[0000-0003-3819-7526]{Raelyn M. Sullivan}\email{rsullivan@phas.ubc.ca}
\author[0000-0003-2730-7419]{Lukas Tobias Hergt}\email{lthergt@phas.ubc.ca}\author[0000-0002-6878-9840]{Douglas Scott}\email{dscott@phas.ubc.ca}
\affiliation{
    Department of Physics and Astronomy,
    University of British Columbia,
    6224 Agricultural Road,
    Vancouver, BC V6T 1Z1, Canada
}

\begin{abstract}
This introductory guide aims to provide insight to new researchers in the field of cosmic microwave background (CMB) map analysis on best practices for several common procedures. We will discuss common map-modifying procedures such as masking, downgrading resolution, the effect of the beam and the pixel window function, and adding white noise. We will explore how these modifications affect the final power spectrum measured from a map. This guide aims to describe the best way to perform each of these procedures, when the different steps and measures should be carried out, and the effects of incorrectly performing or applying any of them. 
\end{abstract}


\section{Introduction}

When studying all-sky cosmic microwave background (CMB) maps, some analysis operations come up frequently, and for non-experts it is not always clear what the optimal choices to make are, and why. As such, this guide is aimed at those who are just starting in the world of all-sky CMB analysis, and those who are more experienced but have wondered whether it {\it really\/} matters if you apodise the mask, smooth by the beam, or apply the pixel window function, or have ever been confused about how to add noise to your map or your power spectrum, or how to downgrade the resolution of your map correctly. We make no claims of completeness but hope this note will act as a useful guide.\footnote{And with details buried in many footnotes!}
This discussion will mostly reference the python package \texttt{healpy} \citep{Zonca2019}, but also the original \texttt{HEALPix} code \citep{2005ApJ...622..759G}.\footnote{Some results in this paper were derived using the \texttt{healpy} and \texttt{HEALPix} packages, \url{http://healpix.sourceforge.net}} We will discuss data sets with a wide range in angular resolution, beam size, and masking. The relevant codes mentioned here are still being updated, so conventions may change, functions may be updated, or new functionality added.\footnote{For example, while writing this document there was a \texttt{HEALPix} version released using Julia exclusively, \url{https://ziotom78.github.io/Healpix.jl/dev/}} It is highly recommended that users also read ``\href{https://arxiv.org/pdf/astro-ph/9905275}{The \texttt{HEALPix} Primer}'' \citep{1999HealpixPrimer}, which covers some of the same topics in greater detail, though with fewer example figures.  

\vspace{0.3cm}
\section{Recommended reading guide for beginners}
Before delving into the details described here, new researchers in the field of CMB analysis should read several other sources. Basic information on the CMB can be found in \cite{ScottSmoot}. For using \texttt{healpy} and \texttt{HEALPix} it is recommended that the user reads \citet{2005ApJ...622..759G} and goes through the \href{https://healpy.readthedocs.io/en/latest/tutorial.html}{tutorials on the \texttt{healpy} documentation}. For results regarding masking it is recommended to read through \citet{2002master}, \citet{2002cutsky} and the documentation by \citet{2023namaster} if using \texttt{NaMaster} to obtain the pseudo-$C_\ell$s. In general, for covering several background topics regarding map-making procedures and noise properties, it is recommended that the user start with the big picture described in \citet{1997tegmark}. 
There are many other useful resources at an accessible level for the new researcher, but starting with these will give good preparation for beginning your own CMB map analysis. 

\vspace{0.3cm}
\section{The basics}

A \texttt{HEALPix} ({\bf H}ierarchical {\bf E}qual {\bf A}rea iso{\bf L}atitude {\bf Pix}elation) map is pixelised such that data on a sphere have pixels of equal area, aligned around lines of equal latitude and with different resolutions nested in a simple way. This is very useful for integration-type procedures and thanks to the availability of the library of routines, \texttt{HEALPix} is very easy to use. 
 
A \texttt{HEALPix} map's resolution is defined by its $N_\mathrm{side}$. The $N_\mathrm{side}$ of the map must be a power of 2 ($1,2,4,16, \dots$) and the number of pixels in a map is then $N_\mathrm{pix}=12\times N_\mathrm{side}^2$, such that an $N_\mathrm{side}=1$ map has 12~pixels, an $N_\mathrm{side}=2$ map has 48~pixels, and so on.\footnote{You can also find the number of pixels using the \texttt{healpy} function \texttt{healpy.nside2npix(nside)}, where \texttt{nside} is the desired $N_\mathrm{side}$.} 
The side-length resolution of a pixel (which we will call $\theta_\mathrm{r}$) of a particular $N_\mathrm{side}$, can be crudely approximated as (in radians)\footnote{\Cref{eq:theta} is a gross approximation, given that the pixels vary in their shapes depending on their location on the sphere. It can also be found using the \texttt{healpy} function \texttt{healpy.nside2resol(nside)}.}
\begin{equation}
    \label{eq:theta}
    \theta_\mathrm{r} = \sqrt{\frac{4\pi}{N_\mathrm{pix}}} = \sqrt{\frac{4\pi}{12 N_\mathrm{side}^2}}\, .
\end{equation}

The pixel data in a \texttt{HEALPix} map may be stored in two possible formats, called `Nested' and `Ring' ordering. The nested choice `nests' the information within large pixels on the sphere, whereas the ring ordering is arranged within rings of data of approximately equal latitude travelling down from the north pole in a spiral. Both are one-dimensional arrays of information, and most \texttt{HEALPix} (and \texttt{healpy} by extension) functions understand both formats. 

The \textit{Planck} \citep{Planck2018I} maps from the Planck Legacy Archive (PLA),\footnote{\url{https://pla.esac.esa.int}} for example, are all stored in ring format, and this is often (but not always) indicated in the FITS\footnote{Flexible Image Transport System (FITS) was designed specifically for astronomical data, but is more generically useful for multi-dimensional arrays or tabular data.} files. For large data runs it is worth testing both in the ring and nested format to compare their efficiencies -- nested is often more parallelisable, but the ring format has been optimised such that it will often run equally fast or faster when accounting for conversion time between ring and nested formats.\footnote{You can convert between the ring format and the nest format using \texttt{healpy.pixelfunc.reorder(map, r2n=True)} and from nest format to ring format using \texttt{healpy.pixelfunc.reorder(map, n2r=True)}.} It is good practice to save maps in a FITS format with all relevant data in the header, including the units for example, in addition to the nest/ring format. 

\vspace{0.3cm}
\section{Going from the power spectrum to a map and back again}
If you are making your own simulated CMB maps you will often start from an idealised power spectrum. You might have obtained this from the best fit from the PLA, or perhaps you produced it using the Boltzmann codes \texttt{CAMB}~\citep{2000Camb,2011camb} or \texttt{CLASS}~\citep{Class1,Class2,Class3,Class4}. The file will include the spectra and cross-spectra, i.e.\ temperature and polarisation \{$TT,EE,BB,TE,TB,EB$\} or perhaps only \{$TT,EE,BB,TE$\}.\footnote{Note that the ordering of the spectra can vary between codes, e.g.\ \texttt{CAMB}'s function \texttt{get\_cmb\_power\_spectra} outputs the spectra in the order diagonals first \{$TT,EE,BB,TE$\}, whereas \texttt{healpy} will by default expect to receive the spectra ordered by row, i.e.\ \{$TT,TE,TB,EE,EB,BB$\} or \{$TT,TE,EE,BB$\}. Since $TB$ and $EB$ are expected to be zero for standard cosmologies they will often be omitted or just arrays of zeros. Newer versions of \texttt{healpy} allow the user to specify the diagonal ordering with the keyword argument \texttt{new=True}.}

For any of $T$, $E$ or $B$, we can write the spherical harmonic transform as\footnote{Sometimes the real-space (or map-space) quantities are distinguished from the harmonic quantities by adding a tilde, e.g.\ $a(\theta,\phi)$ vs $\tilde{a}_{\ell m}$, similarly to how it is often done for Fourier space. We omit such a tilde, here, leaving the designation of harmonic space to a subscript $\ell$ or~$m$.}
\begin{align}
a(\theta, \phi)  &= \sum_{\ell=0}^{\ell_\mathrm{max}} \sum_{m=-\ell}^{\ell} a_{\ell m} Y_{\ell m} (\theta, \phi)\, , \label{eq:a_alm}\\
C_\ell&=\langle |a_{\ell m}|\rangle\, ,\\
a_{\ell m} &= \int_0^{2\pi}\int_0^\pi Y^*_{\ell m}(\theta,\phi) a(\theta, \phi)\sin(\theta)d\theta d\phi\, ,
\end{align}
where $\theta,\phi$ in map space are replaced by discrete pixels\footnote{You can go from $\theta,\phi$ to a pixel index using $n_\mathrm{pix}=$\texttt{hp.ang2pix(\nside,$\theta$,$\phi$)} and from a pixel index to the angular coordinates of the centre of the pixel using \texttt{$(\theta,\phi)=$hp.pix2ang(\nside,$n_\mathrm{pix})$}. There are similar functions to obtain $(x,y,z)$ coordinates as well.}, and the maximum multipole $\ell_\mathrm{max}$ is in principle $\infty$, but in practice for a particular map will be 2 or 3 times $N_\mathrm{side}$, and the minimum is in practice $0$ but usually starts at $\ell=2$ with the first two (for the monopole and dipole) either omitted or set to zero. For the case that the $a_{\ell m}$ are complex rather than real, the sum over $m$ in \cref{eq:a_alm} will run from 0 to $\ell$ (rather than $-\ell$ to $+\ell$).\footnote{\texttt{HEALPix} will handle the various transforms to produce the spectra, cross-spectra and maps, with an extensive spherical transform library.} The conversion from map space to harmonic space is not perfect though; in general, $\ell$ above 2\nside\ should not be trusted unless you dramatically increase the number of iterations
used for the integration (set by the \texttt{iter} parameter in the \texttt{healpy.sphtfunc.map2alm} function call).\footnote{The iterations serve to reduce the errors from the quadrature integration. Each iteration consists of an additional backward and forward harmonic transform. First, a residual map is computed from the ingoing map and the backward transform. Next, the forward transform of this residual map is used to provide a correction to the initial estimate of the harmonic transform. The standard deviation of the residual map is expected to get smaller and smaller with each iteration.}
A better choice is to use \texttt{healpy.sphtfunc.map2alm\_lsq(maps, lmax, mmax, tol=1e-10, maxiter=20)} which will run until either the desired tolerance (\texttt{tol}) is met, or the maximum iterations have passed (\texttt{maxiter}) and will also return a residual map. For a bandlimited map (which is to say a map with little to no signal above $\ell_\mathrm{max}$) you can set the keyword \texttt{use\_pixel\_weights=True} to use the precomputed pixel weights in the integration. For non-bandlimited maps you can smooth with a Gaussian beam to reduce the signal at high $\ell$, or increase the number of iterations and force the results to a higher $\ell_\mathrm{max}$. 
The maximum multipole limit~$\ell_\mathrm{max}$ loosely comes from the intrinsic resolution of the map, since a multipole $\ell$ can be approximated to be a measure of a wavelength on the sphere of $\lambda=2\pi/(\ell+1)$.\footnote{\texttt{healpy} will by default set the limit to be $3\, N_\mathrm{side}-1$ (somewhat arbitrarily), but you can choose a higher or lower $\,\ell_\mathrm{max}$ manually.}  

To generate the $a_{\ell m}$s from the $C_\ell$s in the simplest case, without accounting for the various cross-spectra, each $a_{\ell m}$ will be drawn from a Gaussian distribution with a mean of zero and a variance given by the value of $C_\ell$ at that $\ell$, 
\begin{align}
    a_{\ell m}\sim\mathcal{N}(0,C_\ell)\, ,
\end{align}
which can then be pixelated into a map via \cref{eq:a_alm}.\footnote{To compute a set of $a_{\ell m}$ from $C_\ell$ use the function \texttt{synalm}, to obtain the $a_{\ell m}$s from a map use \texttt{map2alm}, to compute the $C_\ell$s from the $a_{\ell m}$s use \texttt{alm2cl} and to compute the $C_\ell$s from a map use \texttt{anafast}.}
Often the spectra and cross-spectra are plotted as\footnote{This is because $D_\ell$ is effectively the anisotropy power per logarithmic interval in $\ell$ and also a quantity that is approximately constant at low multipoles for a scale-invariant spectrum of perturbations.}
\begin{equation}
    D_\ell \equiv \frac{\ell(\ell+1)}{2\pi}C_\ell ,
\end{equation}
whereas $C_\ell$ is what you use to make the maps. The map units can also vary from CMB temperature (Kelvin, \unit{\kelvin}, or \unit{\micro\kelvin}, also or referred to the CMB blackbody, K$_{\mathrm{CMB}}$\footnote{Often intensity at a given wavelength will be normalised to the equivalent brightness temperature for a \SI{2.7255}{\kelvin} blackbody.}) in terms of fluctuations related to the CMB monopole (i.e.\ dimensionless $\Delta T/T$ units) or Rayleigh-Jeans intensity units (usually for higher frequency channels).

An additional complication is that the non-zero cross-spectra mean that the final $C_\ell$ will need to interrelate the various $T,E$ and $B$ maps. Given some $C_\ell$ for $TT, EE, BB, TE, TB, EB$ the $a_{\ell m}$ will need to correctly account for all the cross-spectra in the calculation of the final $T$, $E$ and $B$ spectra \citep{1999HealpixPrimer,2013Libsharp}. 
In map space, it is more common to measure polarisation using the Stokes $Q$ and $U$ parameters, rather than $E$ and $B$ mode polarisation.  Here $Q$ is the intensity of the radiation polarised in the `plus' orientation (up/down minus left/right, where `up' typically points towards Galactic north), and $U$ is rotated by 45$^\circ$, as a `cross'. Polarisation has an orientation on the sky, compared to the pure intensity or temperature measurements, and as such has a coordinate-system dependence. For more on polarisation, we recommend \citet{1997polPrim}, but also see \citet{1997Kamiokowski} and \citep{1997SeljakZaldarriaga}.

We can go from $Q$ and $U$ maps to $E$ and $B$ power spectra by taking a spin-2 spherical harmonic decomposition of the pseudo-vectors, 
\begin{align}
    (Q+iU)(\theta, \phi)=\sum_{\ell=2}^\infty\sum_{m=-\ell}^\ell a_{\ell m}^{(\pm 2)}{_{\pm 2}}Y_{\ell m}(\theta, \phi)\, .
\end{align}
We can then convert from the spin-2 form to the scalar form,
\begin{align}
    a_{\ell m}^E&=-\frac{1}{2}\left(a_{\ell m}^{(2)}+a_{\ell m}^{(-2)}\right),\\
    a_{\ell m}^{B}&=\phantom{-}\frac{i}{2}\left(a_{\ell m}^{(2)}-a_{\ell m}^{(-2)}\right)\, .
\end{align}

\vspace{0.3cm}
\section{Cosmic variance}
\begin{figure}[tbp]
    \centering
    \includegraphics[width=\textwidth]{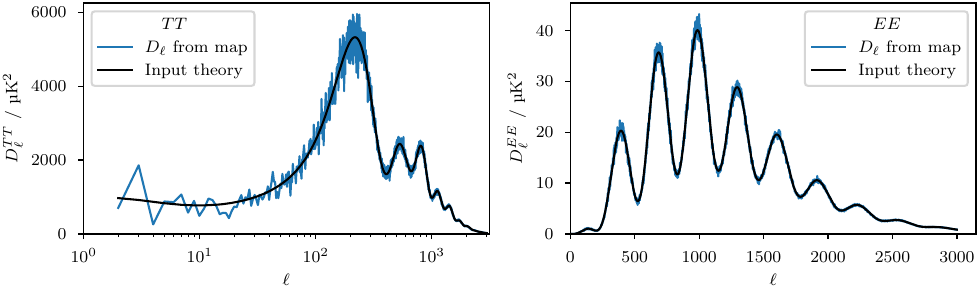}
    \caption{
        Comparison of taking a pure smooth theoretical CMB power spectrum, going to a map, and then computing the power spectrum once more. 
        Note how the scatter is large for the lower multipoles and decreases at higher multipoles. This demonstrates the notion of `cosmic variance', which is to say that we can only measure one sky, so the intrinsic variance on the lowest multipoles is large, since it can only be sampled a small number of times across the sky. A $TT$ example is shown on the left and $EE$ on the right.
    }
    \label{fig:TTClsFromMap}
\end{figure}

Our CMB sky is simply one realisation of the power spectrum, which in the simplest models is Gaussian and random, with the modes chosen to have a mean of zero and a standard deviation given by the power spectrum at each multipole~$\ell$. When you initially take a smooth and ideal CMB power spectrum to make a map, and then convert back to the power spectrum, you will see that the resulting spectrum has a much more scattered appearance than the input (see \cref{fig:TTClsFromMap}). Each time you generate a map you will get a different version of this randomness, but all are statistically identical. This is what is being referred to when we talk about `cosmic variance' -- the notion that since we can only observe a single Universe, there is an intrinsic scatter that exists between the ensemble average and our single sky.\footnote{The origin of the term `cosmic variance' is unclear, but the concept existed in several papers in the 1980s \citep[e.g.][]{AbbottWiseApJL,Fabbri1987,ScaramellaVittorio}.} This scatter can be calculated using
\begin{align}
    \mathrm{Var}[C_\ell^{XY}]=\frac{1}{2\ell+1}\left[C_\ell^{XX}C_\ell^{YY}+\left(C_\ell^{XY}\right)^2\right]\, ,
\end{align}
where $XY\in \{TT,EE,BB,TE,TB,EB\}$ \citep{EisensteinHT,ScottCNM}. This is the lowest theoretically possible variance for the power spectra.\footnote{Looking at cosmic variance in more detail, it arises not only from having only a single universe to observe, but also from only observing a single last scattering surface of the CMB from our location in the Universe. The latter uncertainty could be reduced and thus slightly more information derived by probing other last scattering surfaces through scattering processes of quadrupolar anisotropies \citep{1997Kamionkowski_CV,Adil2024}.} 
Our sky is just one random realisation of the underlying power spectrum, and for the lowest $\ell$ modes we can make very few independent `observations' of these modes ($2\ell+1$ at most) to estimate the power.\footnote{The probability distribution for $C_\ell$ is $\chi^2_k$ with $k=(2\ell+1)$ degrees of freedom.} This cosmic variance is, of course, largest for the smallest multipoles, and decreases with higher $\ell$. When we map only a fraction~$f_\mathrm{sky}$ of the full sky the variance is larger, by a factor of $4\pi/f_\mathrm{sky}$, which is called the `sample variance' \citep{1994scott}. 

\vspace{0.3cm}
\section{The beam}
When a CMB telescope is pointed at the sky, it takes a fuzzy picture. We go from what is an infinite-resolution majesty down to a set of blurry blobs. This is the effect of the beam of the telescope, which averages the sky within its resolution element, with a shape that is approximately Gaussian in the simplest case.
This Gaussian will have some full-width-at-half-maximum (FWHM) size, often quoted as the telescope's angular resolution. Real data processing can be much more complicated, with experiments like \textit{Planck} having different beams \citep{LFIbeams,HFIbeams} for the different frequencies.\footnote{There can even be different beams for different detectors, and complicated scanning patterns can lead to `effective beams', which are the weighted sum of several instantaneous beams.} However, at its most basic level, the telescope's optics consisted of a 2D Gaussian profile convolving our real sky. 
To emulate this when you are building a smoothed map, you should convolve the power spectrum with a Gaussian beam; \texttt{HEALPix} and \texttt{healpy} allow you to do this quite easily, with several Gaussian smoothing options, and options to smooth with a custom beam as well.\footnote{\texttt{HEALPix} by default will convolve with a beam of \SI{420}{\arcmin} when using the routine \texttt{synfast}, whereas \texttt{healpy} will not.} 

\begin{figure}[tbp]
    \centering
    \includegraphics{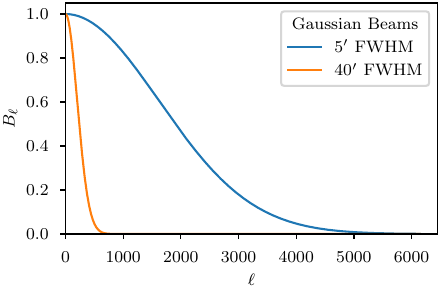}
    \caption{Comparison of beams of FWHM corresponding to \SI{5}{\arcmin} and \SI{40}{\arcmin}. The larger the FWHM the more quickly the higher modes are smoothed or damped out. }
    \label{fig:beamplots}
\end{figure}

\begin{figure*}[tbp]
    \centering
    \includegraphics[width=\textwidth]{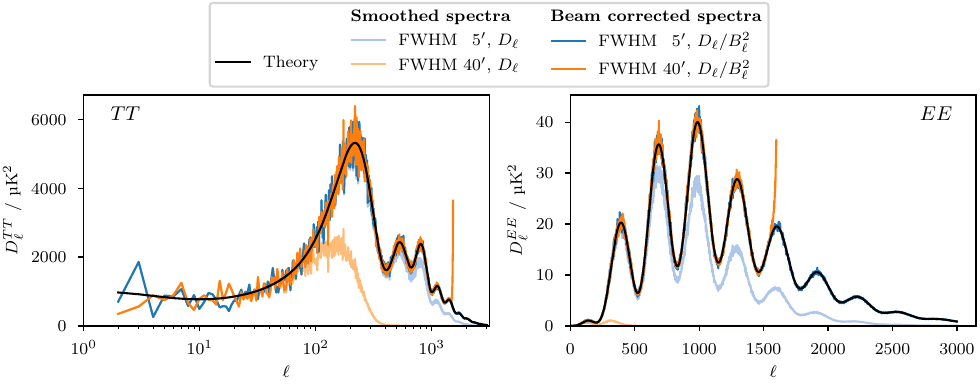}
    \caption{Comparison of power spectrum results from Gaussian-beam-smoothed maps for $TT$ and $EE$ power spectra. Dividing by the beam squared, in this case, returns a good approximation for the input theory. In more realistic cases, with the presence of noise, this division by the beam amplifies the noise at high multipoles. This can also amplify numerical uncertainties at high $\ell$ when smoothing is more extreme, as seen in the spike around $\ell$ of 1500 in the deconvolved $\mathrm{FWHM}=\SI{40}{\arcmin}$ case.
    \label{fig:cls_smoothed}}
\end{figure*}

In the power spectrum the effect of the beam is to damp the power at high multipoles; the smaller your beam the higher $\ell$ you retain good information on, and the bigger the beam the quicker you lose this information (see \cref{fig:beamplots,fig:cls_smoothed}).
To compute the harmonic spectrum profile~$B_\ell$ of a Gaussian beam with standard deviation~$\sigma$:
\begin{align}
    \mathrm{FWHM}&=\sqrt{8\ln(2)}\sigma;\\
    B_\ell&=e^{-\frac{1}{2}\ell(\ell+1)\sigma^2};
    \label{eq:beamT}\\
    _2B_\ell&=e^{-\frac{1}{2}\ell(\ell+1)\sigma^2}e^{2\sigma^2};\label{eq:beamPol}\\
    C_\ell^\mathrm{obs}&=B_\ell^2C_\ell.
    \label{eq:BeamEqus}
\end{align}
The beam function\footnote{Also sometimes written $\exp\{-\frac{1}{2}[(\ell+\frac{1}{2})\sigma]^2\}$ \citep[e.g.][]{ScaramellaVittorio}; if the difference matters to you, then you probably need to use a more accurate beam model in any case.} in \cref{eq:beamT} is the harmonic transform of a Gaussian function on the sky with width $\sigma$.  The factor of $e^{2\sigma^2}$ in \cref{eq:beamPol} is an approximation to go from the temperature or scalar field beam to the beam appropriate for the polarisation (or spin-2) fields \citep{2000Challinor}. This factor is typically small, e.g.\ for a beam with a FWHM of \SI{5}{\arcmin} it is $1+7.63\times10^{-7}$. The beam convolves the sky, and hence \cref{eq:BeamEqus} reminds us that we need the square of $B_\ell$ for the power spectrum.  For precise results when dealing with real experimental data, a measured beam function is usually supplied, which should be used in place of the Gaussian approximation.
To simulate a real experiment, or to compare a real experiment to theory, it is important to include the beam in your considerations (either deconvolving the data with the beam or convolving the beam with the theory). 

\vspace{0.3cm}
\section{The pixel window function}
The pixel window function is an often misunderstood step in the simulation process. What this does \emph{not} do is model the actual pixelisation process performed by \texttt{HEALPix}. What it \emph{does} do is model how observations on the sky are averaged into pixels. In a typical experiment, there will be many observations at many points on the sky and you will average these observations into a single value within each individual pixel. The pixel size is often chosen to have about 3 pixels per beam FWHM.\footnote{This is sometimes referred to as `Nyquist sampling', but unlike in actual Nyquist sampling you \textit{can} gain more information from the CMB power spectrum by using smaller pixels, albeit at the cost of higher noise per pixel and greater computational cost; in other words, this factor of 3 is just a guideline, not a rule. You should always have pixels that are smaller than the beam so that you can sample the beam properly, and at ${>}\,3$ pixels per beam the effect of the pixel window function is small compared to the beam-window function.$^\text{\ref{footnote}}$ This dominance of the beam happens for somewhat smaller pixel-per-beam numbers than you might expect because the effects of the window functions will add in quadrature, and the variance of a box-car function is $(\textrm{width})^2/12$, compared to a Gaussian, which has a variance of $(\textrm{FWHM})^2/(8\ln{2})$.}
 
In any case, you must convolve with the pixel window function to model this averaging step. Depending on your level of interest in higher multipoles, the size of the beam, and the pixelisation chosen, the pixel window function will usually be less significant than the beam (but is still important!).\footnote{We want to emphasize here that even though the pixel window function has a lesser effect than the beam it should \textit{not} be ignored. \label{footnote}} Like the beam, this function can be arbitrarily complicated -- since the telescope's scanning pattern, the shape of the pixels, and the telescope's beam, may all come into play in the final averaging step. Nevertheless, it is often sufficient to use the pixel window function provided by \texttt{HEALPix} and \texttt{healpy} to model this step (although it assumes that your beam is constant across the sky, the scanning pattern is uniform, and that the pixels are well and evenly sampled).

\begin{figure*}[tbp]
    \centering
    \begin{subfigure}
        \centering
        \includegraphics[width=\textwidth]{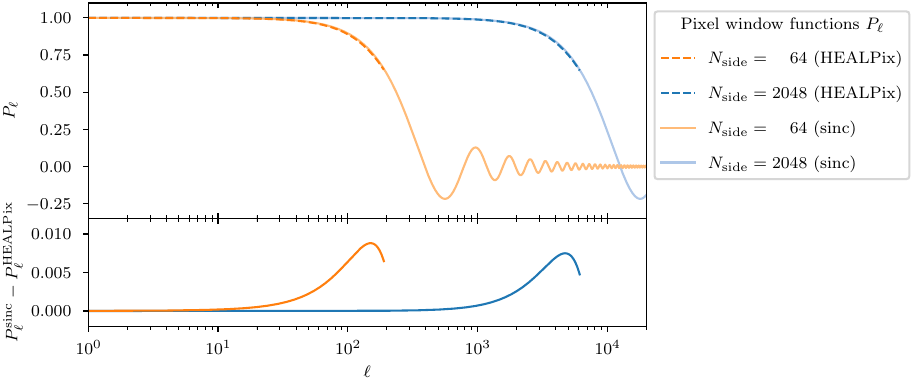}
    \end{subfigure}
    \includegraphics[width=\textwidth]{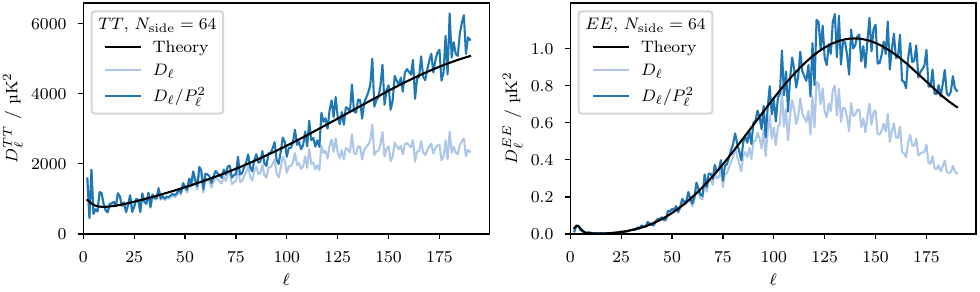}
    \caption{
        Effects of the pixel window function. The top panel shows this function for different $N_\mathrm{side}$ values; 
        the $N_\mathrm{side}=64$ map loses a significant amount of power by $\ell=150$, whereas for $N_\mathrm{side}=2048$ a similar amount of power is lost only by $\ell=5000$. 
        The bottom panels show the resulting power spectra from a map convolved with the pixel window function at $N_\mathrm{side}=64$ and the deconvolved spectrum once divided by the pixel window function squared. 
    }
    \label{fig:pixwin}
\end{figure*}

You can see that the pixel window function is approximately a $\mathrm{sinc}$ function in the top panel of \cref{fig:pixwin}: 
\begin{align}
    P_\ell^\mathrm{sinc}=\mathrm{sinc}\left(\frac{\ell\theta_\mathrm{r}}{2\pi}\right)\, ,
\end{align}
where $\theta_\mathrm{r}$ is the side-length resolution of a pixel defined in \cref{eq:theta}. 
The functions provided by \texttt{HEALPix} are more sophisticated, however, and should be used preferentially over the $\mathrm{sinc}$ approximation.\footnote{\texttt{HEALPix} will by default convolve with the pixel window function, whereas \texttt{healpy} will not when using \texttt{synfast} or \texttt{alm2map}, unless you explicitly set \texttt{pixwin=True} in the function call.}  The \texttt{HEALPix} window functions come from the actual spherical harmonic transform of an average \texttt{HEALPix} pixel for \nside\ less than 128, and are extrapolated to higher \nside\ using a top-hat approximation for the pixels \citep[see][]{1999HealpixPrimer}. 

The beam and pixel window functions are very similar in that each is an observation-driven effect, which reduces the observing power of a particular telescope at small scales (high-$\ell$ modes). However, unlike the beam (which is driven entirely by the telescope's construction) there is more choice in the pixelisation scheme, and hence the pixel window function has more freedom. We focus on \texttt{HEALPix} for this guide, but some studies of relatively small patches of sky will choose to use Cartesian mapping and square pixels. What is important is that both the beam and the pixel window function are well understood and accounted for in your analysis. Putting together the beam and the pixel effects, thus improving our model for a realistic sky data set, we can update \cref{eq:BeamEqus} to
\begin{align}
    \label{eq:P2B2C}
    C_\ell^\mathrm{obs}=P_\ell^2 B_\ell^2 C_\ell\, .
\end{align}
 
\vspace{0.3cm}
\section{Downgrading a map}

It is common to downgrade the resolution of a CMB map; lower-resolution maps will have less noise per pixel, require fewer computing resources, and can be simpler to analyse. The cost is the loss of the higher resolution information (although this may be noise-dominated anyway). This procedure sounds simple enough, but if you are striving for precision, then some subtleties need to be considered. While it may be tempting to use the \texttt{healpy} or \texttt{HEALPix} function \texttt{ud\_grade}, this will convolve your standard scalar CMB map (i.e.\ $T$, $E$ or $B$ map) with the new pixel window function. It is, in essence, averaging all the information from the smaller pixels into the larger pixels and gives the appearance, mathematically, of convolving with the new pixel window function. However, this is an imperfect way to picture this procedure, since the initial pixels may not very densely cover the final pixels, and if the initial map already had a beam or pixel window function applied to it then the final window function would need to account for all of those issues as well. This is perhaps argument enough that the \texttt{ud\_grade} function should not be used to downgrade a typical CMB map. For the non-scalar maps, such as the $Q$ and $U$ polarisation maps, the \texttt{ud\_grade} function should \textit{never} be used, since it fails to correctly average the pseudo-vectors in the $Q$ and $U$ maps, treating them as scalar quantities.

Instead of using \texttt{ud\_grade} when downgrading a map, one should first obtain the $a_{\ell m}$s of the map, and downgrade these in harmonic space. 
The functions used for \texttt{HEALPix} and \texttt{healpy} require that the highest multipoles of the input power spectrum are zero -- for this reason, you must manually set these high $\ell$ to zero or risk there being excess power at small scales in your maps. Typically you want to retain $\ell$ no higher than $\ell_\mathrm{max}=3 N_\mathrm{side}-1$, or conservatively $\ell_\mathrm{max}=2 N_\mathrm{side}$.\footnote{When going directly from $C_\ell$s to maps using \texttt{synfast}, it is not as important to ensure that the high $\ell$ are nulled out, since these are handled properly by the map-building function; however, when going from $a_{\ell m}$s to a map using \texttt{healpy}'s \texttt{alm2map} function, it is crucial.}

\begin{figure*}[tbp]
    \centering
    \includegraphics[width=\textwidth]{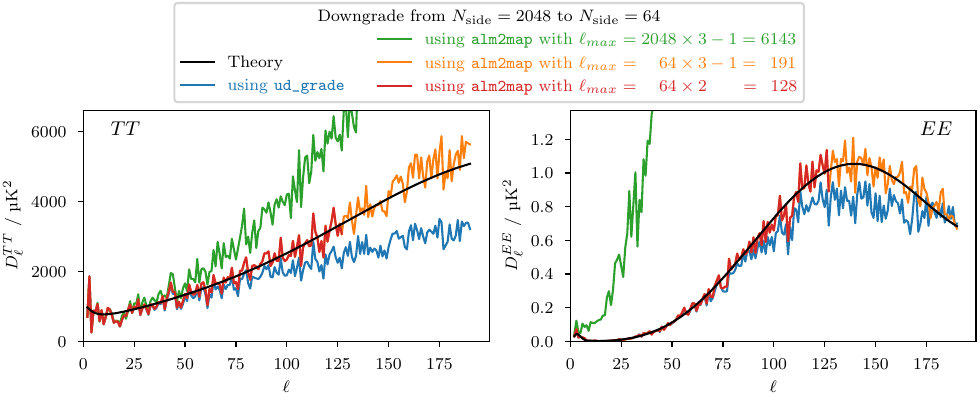}
    \caption{Results for different downgrading methods going from $N_\mathrm{side}=2048$ to $N_\mathrm{side}=64$. The correct downgrading method first takes the map to harmonic space ($a_{\ell m}$), zeros the high-$\ell$ components (above $3\,N_\mathrm{side}-1$, or conservatively above $2\,N_\mathrm{side}$) and finally converts to a map at a lower \nside. In these panels, the orange and red lines show these particular set of procedures. Using the \texttt{healpy.ud\_grade()} function results in low power on small scales for the final maps, represented by the blue line here. This averaging of the pixels has in effect convolved the map with some approximation of the new resolution's pixel window function. The green line shows the results after failing to null the high $\ell$, which leads to an overestimate of the power at small scales -- this can be quite extreme for the $E$ modes especially.
    }
    \label{fig:downgrade}
\end{figure*}

In \cref{fig:downgrade} we show the results from a simple map with no beam and no pixel window function, downgraded from $N_\mathrm{side}=2048$ to $N_\mathrm{side}=64$. First, we downgrade using the \texttt{ud\_grade} function, which results in low power at the highest~$\ell$, due to the implicit convolution with the new pixel window function. This convolution cannot be fixed by simply deconvolving by the pixel window function, since the initial map does not have very good coverage of the final map's pixels. Instead, we take the same initial map and go to $a_{\ell m}$ space. From here we show the results of returning directly back to a lower resolution map, without nulling the high multipoles, and find that it leads to a large excess of power, at the highest~$\ell$ in particular.\footnote{Recall that maps are assumed to have a band-limited signal up to $\ell_\mathrm{max}$; by not nulling the high multipoles the map will not be band-limited to $\ell<3\,N_\mathrm{side}-1$ and there will be excess signal to higher $\ell$.} This can be a substantial effect for the $EE$ power spectrum. Lastly, we show the results after nulling $\ell$ modes above $3\,N_\mathrm{side}-1$ and above $2\,N_\mathrm{side}$ before returning to the lower resolution map; both of these show very good recovery of the input theory and map power spectrum.

In realistic simulations your original map will already have both a beam and a pixel window function; in this case, you must deconvolve with the original functions and reconvolve with the new pixel window function and beam,
\begin{align}
    a_{\ell m}^\mathrm{new}=\left. \frac{B_\ell^\mathrm{new}P_\ell^\mathrm{new}}{B_\ell^\mathrm{old}P_\ell^\mathrm{old}}a_{\ell m}^\mathrm{old} \right\rvert_0^{\ell_\mathrm{max}},
    \label{equ:almsblpl}
\end{align}
where $\ell_\mathrm{max}$ is the new maximum multipole for the lower $N_\mathrm{side}$.
In the case where there are foreground contaminants that you plan to mask later in the analysis, it is a good idea to downgrade the mask similarly. The new mask can be reconverted to simple binary values by choosing a threshold (typically 0.9) to reset the values lower than the threshold to zero, and higher to one. You can choose the threshold to ensure that any foreground contaminants that leak beyond the range of the original mask are still nulled in the new downgraded mask.
If at all possible, you should downgrade the map \emph{before} masking, to ensure that no masking effects are introduced through this process.

\vspace{1cm}
\section{White noise}
The simplest kind of noise to add to the data is Gaussian noise with a white power spectrum, which is flat in harmonic space. 
Further complexities, including the scanning pattern of the telescope, $1/f$ noise, foreground residuals and other details can be added with additional effort, but for the simplest case (as is often desired when trouble-shooting code for example), white noise is a good starting point. 
Adding white noise is a simple procedure and may be done at the map level or the power spectrum level. White noise is a constant addition to each of the multipoles in the power spectrum (in effect simply increasing the standard deviation for each of the modes), or a random number with a mean of zero and a standard deviation at the level of the noise added to each of the pixels in a map. Given the noise~$\noise$ in units of \unit[inter-unit-product=\ensuremath{{}\cdot{}}]{\kelvin\arcmin}, the noise per multipole requires the conversion from arcminutes to radians squared,
\begin{align}
    \noise_\ell = \noise^2 \left(\frac{\pi}{60\times180}\right)^2, 
    \qquad\text{with } 
    [\noise_\ell] &= \unit{\kelvin\squared}, \\
    [\noise] &= \unit[inter-unit-product=\ensuremath{{}\cdot{}}]{\kelvin\arcmin},
\end{align}
and the noise per pixel~$\noise_\mathrm{pix}$ requires division by the side length resolution $\theta_\mathrm{r}$ in arcminutes,
\begin{align}
    \noise_\mathrm{pix} = \frac{\noise}{\theta_\mathrm{r}},
    \qquad\text{with } 
    [\noise_\mathrm{pix}] &= \unit{\kelvin}, \\
    [\noise] &= \unit[inter-unit-product=\ensuremath{{}\cdot{}}]{\kelvin\arcmin}, \vphantom{\frac{\noise}{\theta_\mathrm{r}}}\\
    [\theta_\mathrm{r}] &= \unit{\arcmin}.
\end{align}

As expected, the noise is lower for bigger pixels and larger for smaller pixels and is flat in $\ell$ space. 
As an example, the noise in the \textit{Planck} maps \citep{Planck2018I} was around \SI{2.3e-5}{\kelvin\cdot\arcmin},\footnote{The units $\unit[inter-unit-product=\ensuremath{{}\cdot{}}]{\kelvin\arcmin}$ or $\unit[inter-unit-product=\ensuremath{{}\cdot{}}]{\micro\kelvin\arcmin}$ are often used to describe the sensitivity of a CMB experiment (and particularly to compare experiments), and are sometimes confusing to new researchers. The point is that the per-pixel noise $\noise_\textrm{pix}$ decreases as the square root of the pixel solid angle $\Omega_\textrm{pix}^{1/2}$, and hence the quantity $\noise_\textrm{pix}\theta_\textrm{r}$ is a constant. This is also sometimes referred to as a noise weight per unit solid angle, $w\equiv\noise^2\Omega_\textrm{pix}^{-1}$ \citep{Knox1995}.} so the noise per pixel in an $N_\mathrm{side}=2048$ map would be around \SI{1.3e-5}{\kelvin}, since a pixel has a resolution of \SI{1.72}{\arcmin}, and $\noise_\ell$ would be \SI{4.5e-5}{\micro\kelvin\squared}. 
Often the noise in $Q$ and $U$ maps will be a factor of $\sqrt{2}$ larger than that in $T$, due to the way the data are split, with full data included in the temperature maps, and half the data included in each of the polarisation maps. 

\begin{figure*}[tbp]
    \centering
    \includegraphics[width=\textwidth]{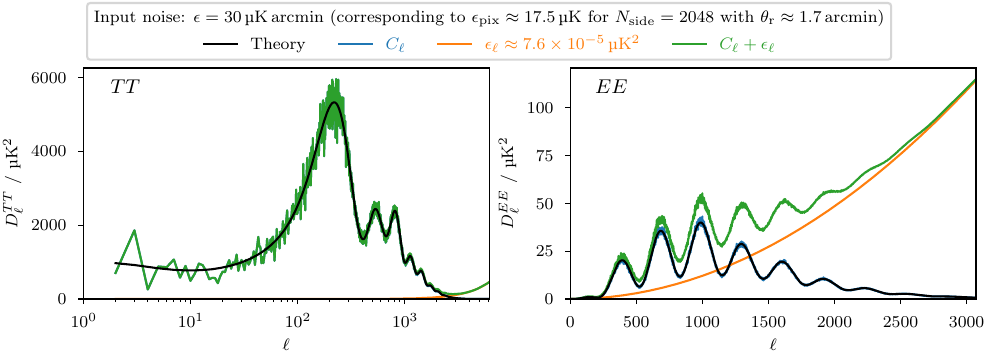}
    \caption{
        Effect of adding white noise to the power spectrum. In this idealistic case, there is no beam or pixel window function and the noise has been added directly to the power spectrum, not the map. 
        The noise level applied here was $\noise_\ell=\SI{7.6e-5}{\micro\kelvin\squared}$ or $\noise=\SI{30}{\micro\kelvin\cdot\arcmin}$. 
        In a more realistic scenario the noise levels for temperature and polarisation would be different, but for simplicity we are assuming the same noise here. 
        Note that since we plot $D_\ell$, the otherwise constant noise spectrum~$\noise_\ell$ increases as $\ell^2$. 
    }
    \label{fig:cls_noise}
\end{figure*}

Depending on how your noise is defined, it generally should not be convolved with the beam and the pixel window function, since this would artificially damp the high-$\ell$ noise (for pure flat white noise). So we can now update \cref{eq:P2B2C} to
\begin{align}
    \label{eq:P2B2CplusN}
    C_\ell^\mathrm{obs}=P_\ell^2B_\ell^2C_\ell+\noise_\ell\, ,
\end{align}
which accounts for the beam and pixel window functions applied to the sky, as well as the white noise. Since $E$~modes are so much weaker, the noise will be a much larger portion of the CMB power spectrum for $E$ modes than for temperature (see \cref{fig:cls_noise}).

\vspace{0.3cm}
\section{The mask}
Of all the procedures described in these notes, masking is likely the most complicated topic that will be discussed. Put simply, the mask\footnote{It is worth pointing out the semantic confusion that sometimes arises here: it is important to be clear whether you are removing or including the data that you `mask'.} should null out areas of the sky that are heavily contaminated with residuals. For the CMB, this is usually the region along the Galactic plane where foregrounds are dominant, but could also involve masking areas around compact sources outside the Galactic plane. 

The simplest mask will consist of ones and zeros, but the sharp edges of the mask can lead to excess power on small scales (high $\ell$). One way to reduce this effect is to soften the mask edges through `apodisation'. 
How you choose to apodise, and by how much, should be carefully considered and tested on any analysis you might perform. Additionally, the method you use to go from the map to the power spectrum and back should be done with care. For high $\ell$ it is preferable to use a pseudo-$C_\ell$ approach, such as the \texttt{MASTER} \citep{2002master} method used by codes like \texttt{NaMaster} \citep{2023namaster}, while for the low $\ell$ it is better to use a quasi-maximum likelihood (QML) method, such as in \texttt{xQML} \citep{2018qml} or \texttt{Eclipse} \citep{2021eclipse,2023eclipse}. There are also methods such as \texttt{PolSpice} \citep{2011polspice} that use the correlation function in map space to estimate the power spectrum. 
We will not recommend the use of any one particular mask, apodisation technique or spherical harmonic decomposition technique here, but rather aim to emphasise some complexities associated with masking. 

\begin{figure*}[tbp]
    \centering
    \includegraphics[width=\textwidth]{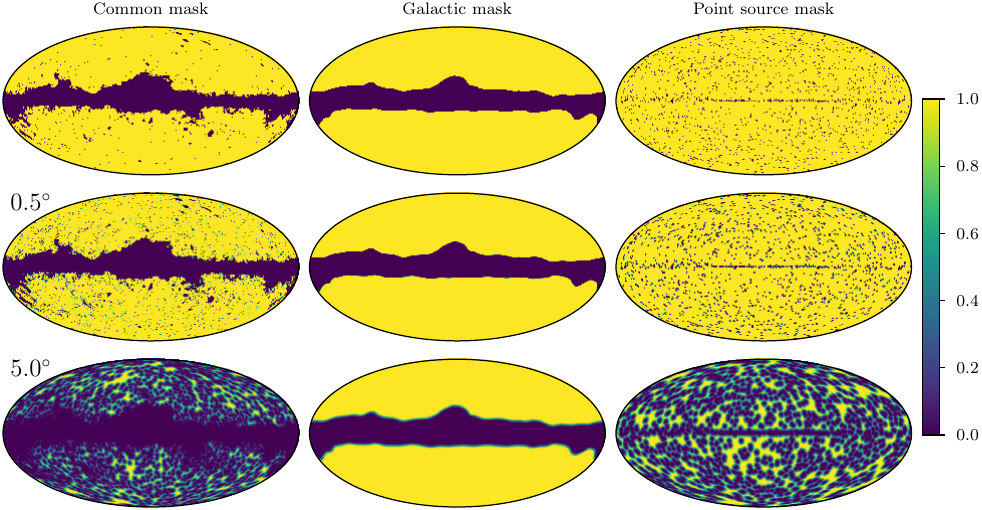}
    \caption{
        Masks compared in this analysis, top to bottom increasing in apodisation length, and left to right the `common' mask (for intensity, there is another common mask specifically for polarisation), the \SI{20}{\percent} Galactic mask (leaving \SI{80}{\percent} of the sky unmasked), and the point-source mask (for the \textit{Planck} lower frequency channels). Yellow corresponds to unmasked regions, whereas dark purple and gradient colours between are masked and apodised regions, respectively. 
    }
    \label{fig:masksused}
\end{figure*}

\begin{figure}[tbp]
    \centering
    \includegraphics[width=\textwidth]{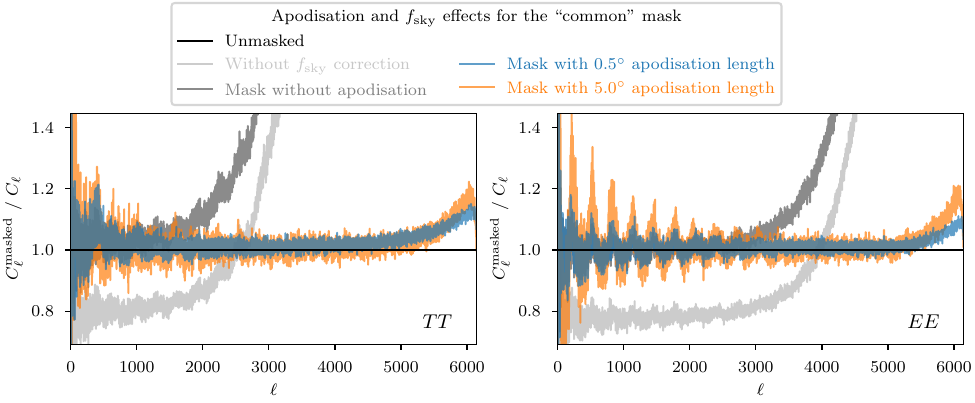}
    \caption{
        Mask apodisation effects. Applying a mask to a CMB map can lead to complicated coupling effects, so obtaining the correct power spectrum from the map must be done with care. 
        Here we show the estimated power~$C_\ell^\mathrm{masked}$ from maps masked with the \textit{Planck} `common' mask relative to the power spectrum~$C_\ell$ from the unmasked map. The left panel shows the results for the temperature spectrum and the right panel the results for $E$-mode polarisation.
        Dividing the observed power by the unmasked sky fraction~$f_\mathrm{sky}$ is an important step in correctly estimating the overall power, but this does not solve the issue that at small scales (high~$\ell$) the power tends to be overestimated.
        Apodising a mask greatly extends the correctly estimated multipole range; however, too extreme an apodisation can eventually degrade the estimate again.
    }
    \label{fig:cls_apodisation}
\end{figure}

To highlight the effects of apodisation (of which \texttt{pymaster}, the Python wrapper of \texttt{NaMaster}, provides a few options), we analyse the impact of the mask on the power spectrum estimate for a range of map types. Specifically, we use the `common' \textit{Planck} mask, a pure Galactic-cut mask, along with a point-source mask. All the masks are provided on the PLA\footnote{\url{https://pla.esac.esa.int/}} in their un-apodised form and for $N_\mathrm{side}=2048$. 
We have used the `C1' apodisation provided by \texttt{NaMaster} on all of these masks, with apodisation lengths of \ang{0.5} and \ang{5.0}; the apodisation procedure smoothly interpolates over that length scale between the `0' and `1' values of the binary masks.\footnote{Note that the \texttt{pymaster} implementation switched the `C1' and `C2' apodisation labels compared to the underlying reference paper by \citet*{Grain2009}.}
\Cref{fig:masksused} gives an overview of all these masks, and \cref{fig:cls_apodisation} shows how apodisation improves the power spectrum estimation, greatly expanding the well recovered multipole range.

Depending on the technique used to estimate the $C_\ell$s, the mask will reduce the amplitude of the power spectrum, which can be corrected by multiplying the theory (or dividing the data) by the fraction of the sky masked,\footnote{The \texttt{HEALPix} and \texttt{healpy} routines \texttt{anafast} will require this correction, but a pseudo-$C_\ell$ estimator, such as \texttt{NaMaster}, generally should not.}
\begin{align}
    f_\mathrm{sky} = \frac{1}{N_\mathrm{pix}} \sum_{i=0}^{N_\mathrm{pix}} M_i^2,
\end{align}
where $M_i=0$ if masked and $M_i=1$ if unmasked for the un-apodised mask, otherwise $M_i$ will be some value ranging from 0 to 1 for unmasked pixels. 
The effect of this $f_\mathrm{sky}$ correction is illustrated in \cref{fig:cls_apodisation}, where the light grey lines correspond to the \texttt{healpy} $C_\ell$ estimates \emph{before} correction, and the dark grey lines to the estimates \emph{after} correction. 
Adding in the masking $f_\mathrm{sky}$ correction, we finally have
\begin{align}
    C_\ell^\mathrm{obs}=f_\mathrm{sky}\left(P_\ell^2B_\ell^2C_\ell+\noise_\ell\right)\, ,
\end{align}
so that we have accounted for the beam, pixel window function, noise, and power reduction due to the mask. 

\begin{figure}[tbp]
    \centering
    \includegraphics[width=\textwidth]{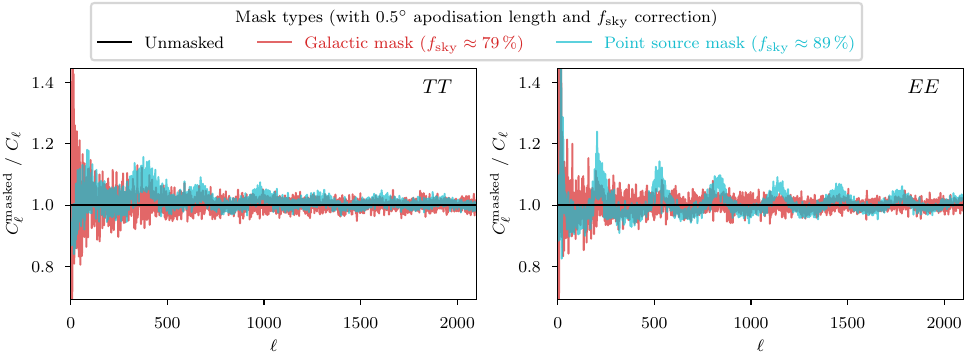}
    \caption{
        Same plot as in \cref{fig:cls_apodisation}, but now highlighting the effect of different mask types, specifically the difference between a Galactic mask in the form of a smooth belt around the equator and a point-source mask in the form of many disconnected circular spots (see \cref{fig:masksused} for a visual comparison).
        Both masks were apodised before power estimation, and both resulting $C_\ell^\mathrm{masked}$ estimates were corrected with the corresponding sky fraction.
        Overall the power estimates are close to the expected values, but there is a clear ringing effect in the results with the point-source mask.
    }
    \label{fig:cls_mask_types}
\end{figure}

\begin{figure}[tbp]
    \centering
    \includegraphics[width=\textwidth]{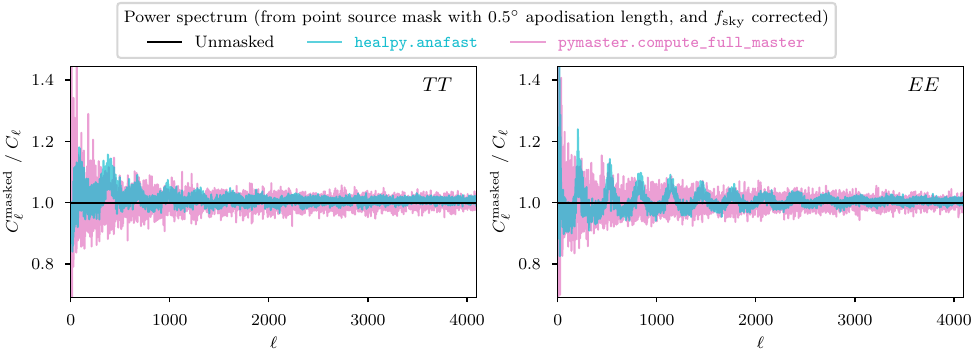}
    \caption{
        Same plot as in \cref{fig:cls_apodisation}, but now highlighting two different power spectrum estimation functions, one from \texttt{healpy} and the other from \texttt{pymaster}.
        The latter manages to recover the unmasked $C_\ell$ slightly better, and visibly reduces the ringing effect from the point-source mask.
    }
    \label{fig:pymaster}
\end{figure}

Apodisation and $f_\mathrm{sky}$ correction might not be enough, though. Some mask types come with unique challenges, e.g.\ the point-source mask introduces heavy ringing in \texttt{healpy}'s power spectrum estimate, as demonstrated in \cref{fig:cls_mask_types}. This ringing effect can be mitigated by switching to \texttt{pymaster} (see \cref{fig:pymaster}), but it comes at the cost of overall larger uncertainty in the estimate.

\vspace{0.7cm}
\section{Final steps and summary}
Let us summarise by listing the steps for generating a simple CMB map realisation with white noise. 
\begin{enumerate}[beginpenalty=10000]
    \item Generate from \texttt{CAMB} or \texttt{CLASS} the desired power-spectrum~$C_\ell$ (or download a pre-computed one, e.g.\ from the PLA).
    \item Convolve the power spectrum with the beam and pixel window functions, $B_\ell$ and $P_\ell$.
    \item Add the desired noise spectrum~$\noise_\ell$.
    \item Generate the map (from the randomly sampled $a_{\ell m}$s).
    \item Mask the map (ensuring that the mask is apodised for certain procedures).
\end{enumerate}
From the map (before masking), there are a few possible processing steps, such as decreasing the map resolution, which we now list. 
\begin{enumerate}[beginpenalty=10000]
    \item Starting from the input map, obtain the $a_{\ell m}$s from the map and the mask.
    \item Cut all the $\ell$ above $2$\nside\ (conservatively) or 3\nside$-1$ (less conservatively). 
    \item Deconvolve by the beam and pixel window functions and reconvolve by the new beam and the new pixel window function.
    \item Downgrade the mask in the same way as the map. The new mask will be extended to account for regions of the sky that were originally masked leaking into unmasked regions. Choose some threshold for this step and re-apodise the mask. 
    \item Generate the new map at the lower resolution, and apply the new downgraded mask. 
\end{enumerate}
Returning to the power spectrum, these are a few more possible steps. 
\begin{enumerate}[beginpenalty=10000]
    \item From a masked map you should first remove the dipole and the monopole (these are usually masking effects and will alter the effectiveness of the final results).
    \item Ensure that the applied mask was apodised, or apply an apodised mask to the map.
    \item Obtain the low-order $C_\ell$s using QML or \texttt{anafast} and the higher orders using a pseudo-$C_\ell$ method such as \texttt{NaMaster} to control for the effects of the mask. If \texttt{anafast} is used throughout then you must divide by the sky fraction to obtain the correct amplitude of the underlying $C_\ell$s. 
    \item Divide the $C_\ell$s by the beam and pixel window functions squared. The noise will dominate over the signal at the high~$\ell$. 
\end{enumerate}

To summarise, going from a CMB map to a power spectrum and back should be done with care, but following the correct procedures will introduce relatively few artefacts, allowing for effective data analysis to be performed in either map space or harmonic space. Any steps taken should be well understood and tested, and any results should be robust against choices made, such as resolution, beams, masks used, and the methods used to go from map to harmonic space and back again. We hope this guide has provided some insights and will help future cosmologists avoid some of the common pitfalls and misunderstandings regarding CMB map analysis.

\vspace{0.3cm}
\section{Acknowledgements}
We would like to thank colleagues who sent us comments and suggestions on an earlier version of this guide.  This work was supported by the Natural Sciences and Engineering Research Council of Canada.
LTH was supported by a Killam Postdoctoral Fellowship and a CITA National Fellowship.
This paper made use of \texttt{astropy},\footnote{\url{https://www.astropy.org}} \texttt{CAMB},\footnote{\url{https://github.com/cmbant/CAMB}} \texttt{HEALPix},\footnote{\url{https://healpix.sourceforge.io}} \texttt{healpy,\footnote{\url{https://github.com/healpy/healpy}}} and \texttt{NaMaster}.\footnote{\url{https://github.com/LSSTDESC/NaMaster}}
\bibliography{bibliography}{}

\bibliographystyle{aasjournal}

\end{document}